\documentclass[twocolumn,a4paper,nofootinbib,
showpacs,aps,floatfix]{revtex4}

\usepackage{graphicx}
\usepackage{bm}
\usepackage{amsmath}
\def\la{\langle}
\def\ra{\rangle}

\begin {document}
\title{Transversal laser excitation of atoms in a waveguide}
\author{I. Lizuain}
\email[Email address: ]{qfblilii@lg.ehu.es}
\affiliation{Departamento de Qu\'\i mica-F\'\i sica,
Universidad del Pa\'\i s Vasco, Apdo. 644, Bilbao, Spain}
\author{A. Ruschhaupt}
\email[Email address: ]{a.ruschhaupt@tu-bs.de}
\affiliation{Institut f\"ur Mathematische Physik, TU Braunschweig, Mendelssohnstrasse 3, 38106 Braunschweig, Germany} 
\author{J. G. Muga}
\email[Email address: ]{jg.muga@ehu.es} 
\affiliation{Departamento de Qu\'\i mica-F\'\i sica,
Universidad del Pa\'\i s Vasco, Apdo. 644, Bilbao, Spain} 


\begin{abstract}
We investigate the effect of a laser beam  perpendicular to 
a waveguide which channels two-level atoms.  
For weak transversal coupling and the laser on resonance with the 
internal atomic transition the excitation of transverse
atomic levels occurs at avoided crossings associated with ``Rabi resonances'' in which the Rabi frequency  
coincides with the transition frequency between the transverse levels.       
\end{abstract}
\maketitle
%
%
%
%
%
\section{Introduction}
The aim of atom optics is to study and control 
coherent atomic waves  
interacting with electromagnetic fields or material structures. 
An important tool to implement this control is   
the spatial confinement of laser cooled atoms, which      
makes possible  
their manipulation, trapping, and transport.  
Reabsorption of scattered photons, in particular,
is reduced, so that quantum coherence is more likely to survive
\cite{CCL98}.    
Waveguides may lead to an atom-optical analog of integrated circuits
and much work is in progress to create atom-optical analogs of
standard optical components for waveguide structures.
Possible applications are information processing and interferometry.   
Waveguides are also interesting because they counteract gravitational acceleration prolonging observation times, and may help to dispense precise quantities  
of atoms onto specific regions of a surface in atom lithography \cite{Key00}. 
There is in addition a growing interest in systems
with low dimensionality because of a wealth of characteristic  
phenomena which do not occur in unconstrained space \cite{HBH98}: 
as an example, waveguides with high transverse frequencies have been proposed to   
realize  
a one dimensional (1D) gas of impenetrable bosons, the so-called Tonks-Girardeau
gas \cite{Tonks}. Moreover, effective 1D guides allow us to examine in simple geometrical configurations
and with the aid of elementary theoretical treatments many fundamental questions
in quantum mechanics such as decoherence, definitions and measurements of quantum times, e.g. the arrival time \cite{TOA1,TOA2}, 
or quantum atom statistics \cite{box}.    

All these recent trends and prospects have motivated experimental and theoretical research on 
localized quantum waves propagating through different 
structures and confinements. 
Several techniques and physical interactions have been used to restrict and 
steer the atomic motion.  
Atom guiding may be simply carried out by dipolar forces in laser standing waves
\cite{channeling}, or along a red detuned laser beam.    
Olshanii et. al. proposed to guide atoms in a red detuned laser beam, itself guided in 
a hollow optical fiber
\cite{Olshanii93} to facilitate bending. This idea has been realized \cite{JILA}.
Another approach is based on 
reflecting the atoms with the evanescent field of blue detuned light modes guided in a glass fiber tube
\cite{MSZR94}.  
Guides based on hollow (doughnut) blue detuned laser beams have been implemented too \cite{Kuga97,Xu99}. 
Other major way of guiding atoms is by means of magnetic fields \cite{HH99, Folman02}, 
in wire-lined hollow fibers or along current carrying wires. 
In addition, beam splitters for guided atoms have been demonstrated \cite{BS00}, as well as waveguides in two dimensions using magnetic \cite{HBH98},
or optical fields \cite{Pfau98}.      

Many intended applications of atoms in waveguides rely on a single-mode   
atomic propagation. In this respect transversal excitations due to waveguide narrowing \cite{Stenholm02}, 
inhomogeneities because of geometric deformations of the current-carrying wires \cite{BECs02}, sudden potential variations \cite{Koehler05}, 
waveguide splitting \cite{Stenholm03}, or bending \cite{bending04}, have been examined, 
as well as excited mode measurement by free expansion \cite{Stenholm02_01}, and the decoherence 
due to thermal fluctuations of the environment \cite{deco03,deco04}. 

\begin{figure}
\includegraphics[height=3cm]{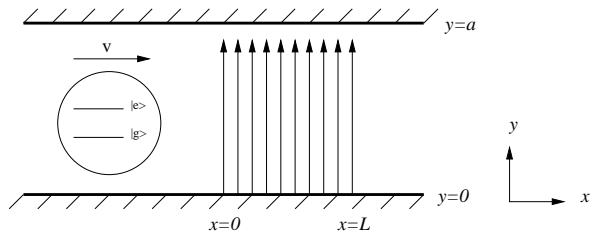}
\caption[]{Schematic representation: a two-level atom, confined in a waveguide
between $y=0$ and $y=a$, impinging with velocity $v$ on a laser illuminated region
between $x=0$ and $x=L$. The atom-laser interaction will in general change the 
internal, transversal, and longitudinal atomic state.}
\label{scheme}
\vspace{1 cm}
\end{figure}

In this paper we start to investigate 
the effects of including a further control element in the waveguide, 
namely, a  
laser beam perpendicular to the longitudinal direction of the guide. 
Such non-guiding fields could be used for many different purposes, such 
as preparation of specific internal, transversal, or translational states, 
including in particular transversal cooling, 
state-selective detection, switching, time dependent trapping, 
or filtering a privileged direction of motion with an ``atom diode''
\cite{AD1,raizen1,raizen2,AD2,quenching}, 
which has been proposed to implement a new cooling method for longitudinal motion \cite{raizen1,raizen2}.   
We shall restrict the present article to a simple model consisting 
in two-level atoms on-resonance with a ``square'' (semiclassical) laser beam 
of finite width $L$, 
neglecting decay and assuming a rectangular, hard-wall waveguide.   
This idealized setting, schematically represented in Fig. \ref{scheme}, has turned out to be interesting and rich enough for a separate and detailed consideration, and additional complications such as 
detuning, decay, or more realistic profiles and confining potentials will be treated elsewhere.
As a main result, it is found that transversal excitation occurs at ``transversal Rabi resonances'' associated with 
avoided crossings of the mode levels with respect to the Rabi frequency.     

There is no direct precedent to this analysis, to the best of our knowledge, but some related works are worth mentioning. In a recent study the possibility to detect 
single, guided atoms crossing a microcavity has been examined, with a classical 
treatment of the atomic motion \cite{SAD03}. Technically our model is the extension to a waveguide configuration of a series of previous one-dimensional models in which simple laser profiles interacting with two or three level atoms have been used to investigate arrival times \cite{TOA1,TOA2}, detection optimization \cite{OD},  Rabi oscillation suppression \cite{ROS}, or ``atom diodes'' \cite{AD1,AD2} 
using a quantum treatment of the translational motion. Indeed the present results will, as we shall see, justify the use of these one dimensional models for small values of the 
ratio between the transverse confinement and the laser wavelength, i.e., for a small 
Lamb-Dicke parameter. This work is also related to the laser induced vibrational transitions in ion-traps, see \cite{IT2,IT3} and references therein.
In that context though, the emphasis is on 
the effect of laser detuning and small Rabi frequencies, whereas we shall restrict the present paper to on-resonance 
(zero detuning) 
interactions. Since the approximations are different (we will not make use of a ``second'' rotating wave approximation with respect to transversal motion), the transitions 
described here do not appear in the usual treatments of ion-traps, but they become very important, as we shall see, at the Rabi resonances mentioned above.         

In Section II the model is described and the corresponding stationary Schr\"odinger equation solved. In Section III 
the main results about transversal excitation are discussed and the paper ends with a discussion and the main conclusions in Section IV. 

%
%
%
%
%
%
%
%
\section{Basic Theory}
\subsection{The Model}
The Hamiltonian describing a two level atom interacting with a classical 
electric field $\boldsymbol{E}(\hat{\boldsymbol{r}},t)=
\mathbf{E_0} \cos {(\omega_L t-\boldsymbol{k_L}\cdot\hat{\boldsymbol{r}})}$
with constant $\mathbf{E_0}$, in the dipole approximation and neglecting spontaneous emission is 
$$
H=\frac{\boldsymbol{\hat{p}}^2}{2m}+H_A+e\boldsymbol{D} \cdot \boldsymbol{E}(\hat{\boldsymbol{r}},t),
$$
%
where 
$$
H_A=\hbar \omega_g |g\rangle\langle g|+ \hbar \omega_e |e\rangle\langle e|
$$
%
is the internal Hamiltonian of the atom, $\boldsymbol{D}$ 
is the atomic dipole operator,
%
$$
\boldsymbol{D}=\boldsymbol{d}|g \rangle\langle e|+ H.c , 
$$
%
$\boldsymbol{d}$ being the transition dipole moment between the states $|g\ra$ and $|e\ra$, and 
$\boldsymbol{\hat{r}}$ and $\boldsymbol{\hat{p}}$ are the atomic center of mass position and momentum operators. 
Taking $\boldsymbol{d}$ real and $\omega_g=0$ for simplicity, 
the Hamiltonian reads
\begin{eqnarray}
H&=&\frac{\boldsymbol{\hat{p}}^2}{2m}+\hbar \omega_e |e\rangle\langle e|
 +e\boldsymbol{d} \cdot \boldsymbol{E_0} \left\{ |g\rangle\langle e|+ |e\rangle\langle g|\right\}
\nonumber\\
&\times&
\cos {(\omega_Lt-\boldsymbol{k_L}\cdot\hat{\boldsymbol{r}})}.
\nonumber
\end{eqnarray}
On resonance, the laser frequency $\omega_L$ is equal to the atomic transition frequency ($\omega_L=\omega_e$). In the interaction picture with respect to
$H_0=\hbar \omega _L | e\rangle\langle e|$, 
and in the usual rotating wave approximation, 
there results a  time-independent Hamiltonian,
%
$$
H_I=\frac{\boldsymbol{\hat{p}}^2}{2m} 
+ 
\frac{\hbar\Omega}{2}\lbrace|g\rangle\langle e|e^{-i\boldsymbol{k_L}\cdot\hat{\boldsymbol{r}}}+
|e\rangle\langle g|e^{i\boldsymbol{k_L}\cdot\hat{\boldsymbol{r}}}\rbrace,
$$
where $\Omega=e\boldsymbol{d\cdot E_0}/\hbar$ 
is the Rabi frequency.
%
%

From now on we shall assume that the laser illuminates only the $0\le x\le L$ region in the
$y$ direction, see Fig. \ref{scheme}.  Since the
Hamiltonian does not depend on $z$, there is free motion in the $z$-direction 
and we shall only consider explicitly the two-dimensional $x$-$y$ plane.
(Alternatively, there may be an additional hard wall confinement in $z$-direction
so that the wave remains in the corresponding ground state.)  
In a representation in which  
$|g\rangle \equiv \left(\begin{array}{c}1\\0\end{array}\right) $ and
$|e\rangle \equiv \left(\begin{array}{c}0\\1\end{array}\right)$, 
the Hamiltonian becomes, in matrix form,
$$
H_I=\frac{\hat{p}_x^2}{2m}+\frac{\hat{p}_y^2}{2m}+
\Theta(\hat{x})\Theta(L-\hat{x})\frac{\hbar\Omega}{2}
\left( \begin{array}{ccc}
0 & e^{-ik_L\hat{y}}\\
e^{ik_L\hat{y}} & 0
\end{array}\right). 
$$
%
%
\subsection{Stationary Solutions}
\label{stationary_solutions}
We now look for 
solutions of the stationary Schr\"odinger equation  
\begin{equation}
\label{schr}
H_I|\psi\rangle=E|\psi\rangle,
\end{equation}
where $E$ is the total energy of the 
incident atom in the interaction picture and the wave function $\psi(x,y)$
is subject to the boundary conditions imposed by the infinitely high walls at
$y=0$ and $y=a$, namely $\psi(x,0)=\psi(x,a)=0$.

Let us denote as $|\varphi_n\rangle$ the normalized energy eigenstates of an 
infinite well in the $y$ direction, 
%
$$
\varphi_n(y)=\langle{y}|\varphi_n\rangle=\sqrt{\frac{2}{a}}
\sin\left({\frac{n\pi}{a}y}\right), \quad n=1,2,...
$$
with energies  
%
$$
E_n=\frac{\hbar^2}{2m}\left( \frac{n\pi}{a}\right) ^2.
$$
Each of these eigenstates is associated with two \emph{quasi-momenta} $\pm p_n=\pm\hbar\frac{n\pi}{a}$, 
which will be useful later on for an intuitive understanding of 
laser coupling of transversal modes. 

\begin{figure}
\includegraphics[height=6cm]{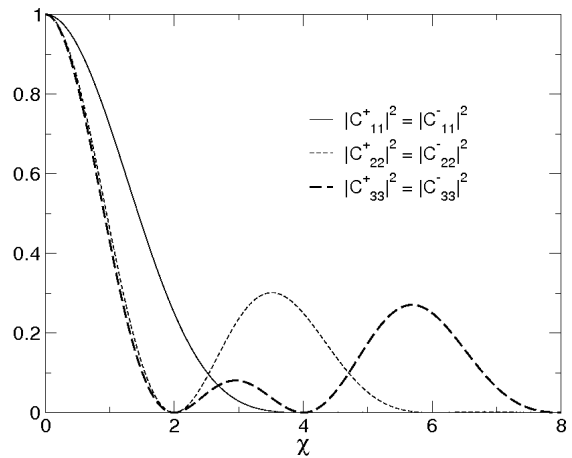}
\caption[]{Diagonal terms of the coupling matrix $\hat{C}$.}
\label{diag_coup}
\vspace{1 cm}
\end{figure}

\begin{figure}
\includegraphics[height=6cm]{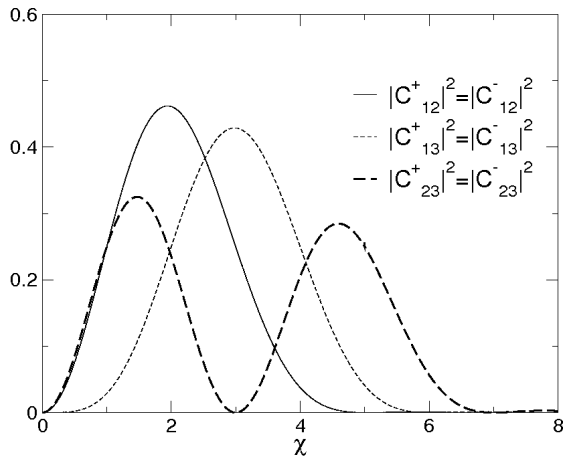}
\caption[]{Off-diagonal terms of the coupling matrix $\hat{C}$ between
the first few modes.}
\label{off_diag_coup}
\end{figure}

We now expand $|\psi\rangle$ in the $\{|x,j,\varphi_n\rangle\}$ basis. The $j$ stands 
for the internal atomic state, i.e., 
$\left(j=g,e\right)$,  
and $\varphi_n$ for the ``free'' transversal mode (``free'' meaning hereafter 
``without laser interaction'', but constrained by the waveguide). 
In principle,  an infinite number of transversal modes
should be considered, $n=1,..., \infty$, 
\begin{equation}
\label{expansion}
|\psi\rangle=\sum_{j=g,e}\sum_{n=1}^{\infty}\int_{-\infty}^{\infty}dx\,|x,j,\varphi_n\rangle\langle x,j,\varphi_n|\psi\rangle.
\end{equation}
As the sum over $n$ in  Eq. (\ref{expansion}) has an infinite number of terms, truncation up
to some finite number of (free) transversal modes $N$ becomes necessary to
perform 
numerical calculations. We will study later the convergence with respect to 
$N$.
Inserting this expansion into the stationary Schr\"odinger equation (\ref{schr}), 
we get for the $x$-dependent amplitudes
\begin{equation}
\label{x_anpl}
\psi_{jn}(x)=\langle x,j,\varphi_n|\psi\rangle
\end{equation}
the equation
\begin{equation}
\label{xequation}
\frac{\hat{p}_x^2}{2m} |\psi(x)\rangle +
\hat{H}_\perp|\psi(x)\rangle+
\hat W|\psi(x)\rangle=
E|\psi(x)\rangle,
\end{equation}
where we have collected all the $\psi_{jn}(x)$ in a  
column $2N$-vector $|\psi(x)\rangle$,
\begin{equation}
\label{vector_1}
|\psi(x)\rangle=
\left( \begin{array}{c}
\psi_{g,1}(x)\\
\vdots\\
\psi_{g,N}(x)\\
\psi_{e,1}(x)\\
\vdots\\
\psi_{e,N}(x)\\
\end{array}\right), 
\end{equation}
$\hat{H}_\perp$ is a diagonal matrix 
with the transversal energies, 
%
$$
\hat{H}_\perp=
\left( \begin{array}{c|c}
\begin{array}{ccc}
E_1 &  &\\
& \ddots & \\
& & E_N\\
\end{array}& 0\\
\hline
0& \begin{array}{ccc}
E_1 &  &\\
& \ddots & \\
& & E_N\\
\end{array}\end{array}
\right),
$$
and the coupling potential $\hat W$ is given by 
%
$$
\hat W=\frac{\hbar\Omega}{2}
\left( 
\begin{array}{c|c}
0 & \hat{C}^{-}\\
\hline
\hat{C}^{+} & 0
\end{array}\right),
$$
%
where the $\hat{C}^{\pm}$ are two $N\times N$ dimensional matrices depending on $k_La$
with elements 
\begin{equation}
\label{coupling_elements_1}
C_{nn'}^{\pm}=\langle\varphi_n|e^{\pm ik_Ly}\varphi_{n'}\rangle=\int_0^a dy \varphi_n(y) \varphi_{n'}(y) e^{\pm ik_Ly}
\end{equation}
%
%
and $n$,$n'=1,\ldots,N$.
The explicit expressions for these coupling elements are
\begin{eqnarray}
C_{nn'}^{\pm} &= & \pm iak_L \left[ 1-e^{\pm iak_L} \left( -1 \right) ^{n+n'} \right]
\nonumber\\
&\times&
\left(
\frac{1}{a^2k_L^2-(n'-n)^2\pi^2}\mp\frac{1}{a^2k_L^2-(n'+n)^2\pi^2}
\right).
\nonumber
\end{eqnarray}
Note that $C_{nn'}^{\pm}=C_{n'n}^{\pm}$ and $\left(C_{nn'}^{\pm}\right)^{\ast}=C_{nn'}^{\mp}$.
The coupling elements for the first few modes are plotted in
Figs. \ref{diag_coup} and \ref{off_diag_coup} versus $\chi:=k_La/\pi$. The maxima 
of these functions will be close to  
the  ``resonance'' conditions, 
\begin{eqnarray}
\chi &=& n'-n,\;\;n'\ge n,
\nonumber\\
\chi &=& n+n',
\nonumber
\end{eqnarray}
namely to laser kicks coinciding with the 
momentum jumps between quasi-momenta
of the transversal modes. 
In fact the two peaks may merge into one if they are close enough,
see e.g. a single peak 
of $C^{\pm}_{13}$ at $\chi=3$ in Fig. \ref{off_diag_coup}. 
%
  
\subsubsection{Solutions in the laser-free region $(x\le 0$ and $x\ge L)$}
In the laser-free region we have that
\begin{equation}
\label{laser_free}
\frac{\hat{p}_x^2}{2m}{\psi}_{j,n}(x)=\left ( E-E_n \right){\psi}_{j,n}(x),
\end{equation}
where $E_{n}$ is the  energy of the $n^{th}$ transversal state.
%
%
%
The solution to Eq. (\ref{laser_free}) in the $x\le0$ region 
corresponding to the incidence of a ground-state atom with a definite positive longitudinal velocity 
in the waveguide's first (ground) transversal mode is given by
%
%
\begin{equation}
\psi_{j,n}(x\le 0)=\delta_{jg}\delta_{n1}e^{ik_n x}+R_{j,n}e^{-ik_n x},
\end{equation}
and for the transmitted part to the right of the barrier,
\begin{equation}
\psi_{j,n}(x\ge L)=T_{j,n}e^{ik_nx},
\end{equation}
where the longitudinal wavenumbers $k_n$ are obtained from  
\begin{equation}
\label{k_alpha}
k_{n}^2=\frac{2m}{\hbar^2}\left(E-E_{n}\right)
\end{equation}
%
taking positive roots for $k_n^2>0$ and positive imaginary 
roots for evanescent closed channels $(k_n^2<0)$. 
The, so far unknown, reflection and transmission amplitudes, namely 
$R_{j,n}$ and $T_{j,n}$,
will be determined from the matching conditions
at $x=0$ and $x=L$.
Evaluating from these solutions incoming and outgoing fluxes,
the probabilities for the different outgoing channels satisfy
\begin{equation}
\label{probabilities}
1=\sum_{j=g,e}\sum_n |R_{j,n}|^2\frac{k_n}{k_1}+
\sum_{j=g,e}\sum_n |T_{j,n}|^2\frac{k_n}{k_1},
\end{equation}
where the sum extends over asymptotically open channels
with $k_n$ real, since the imaginary ones
do not contribute to the asymptotic flux.  

%
%
%
\subsubsection{Solutions inside the laser $(0\le x\le L)$}
The equation to solve inside the laser-illuminated region is
\begin{equation}
\label{inside_laser}
\frac{\hat{p}_x^2}{2m}|\psi(x)\rangle+\hat{H}_\perp|\psi(x)\rangle+
\hat W|\psi(x)\rangle=E|\psi(x)\rangle.
\end{equation}
We denote by $\hat {\cal{E}}$ the following operator,
\begin{equation}
\label{a_definition}
\hat {\cal{E}}= E\boldsymbol{1}_{2N}- \hat{H}_\perp-\hat W.
\end{equation}
Since the (longitudinal) kinetic part $\frac{\hat{p}_x^2}{2m}$ of the
Hamiltonian  commutes with $\hat{\cal{E}}$, we can separate variables,  
diagonalizing the ``non-kinetic part'' and multiplying the resulting
eigenstates  
(which represent the natural modes in the laser-illuminated region) by 
plane waves having the remaining energy.   
Therefore, if $|\epsilon_{\alpha}\rangle$ is the $\alpha^{th}$ eigenvector of
the $\hat{\cal{E}}$ matrix
and $\epsilon_{\alpha}$ is the corresponding eigenvalue, the solution in the laser region
will be a linear combination of the form 
\begin{equation}
\label{gral_sol_x}
|\psi(0\le x \le L)\ra=\sum_{\alpha=1}^{2N} \left( A_{\alpha}|\epsilon_{\alpha}\rangle e^{iq_{\alpha}x} +
 B_{\alpha}|\epsilon_{\alpha}\rangle e^{-iq_{\alpha}x} \right)
\end{equation}
%
%
where ${q^2_{\alpha}}=\frac{2m\epsilon_{\alpha}}{\hbar^2}$ ($q_\alpha$ is 
the positive root for $\epsilon_\alpha\ge 0$ and positive imaginary otherwise,
to represent outgoing or evanescent waves respectively). 
The continuity of the the wave function $\psi(x,y)$ and its derivative $\psi'(x,y)$ 
at $x=0$ and $x=L$ for all $y$, leads to a linear system of $8N$ equations and $8N$ 
unknowns, from which all the reflection, transmission and eigenmode amplitudes $A_\alpha$ and $B_\alpha$ can be obtained.
%
For each mode $|\epsilon_\alpha\rangle$ in the laser region, 
$q_\alpha$ is the corresponding longitudinal wavenumber, and $\epsilon_\alpha$ the longitudinal translational energy. 
Notice that, at variance with the laser free region, the eigenmodes with laser interaction imply in general 
a non-factorized combination of internal and (free) transversal states. 
\begin{figure}
\includegraphics[height=6cm]{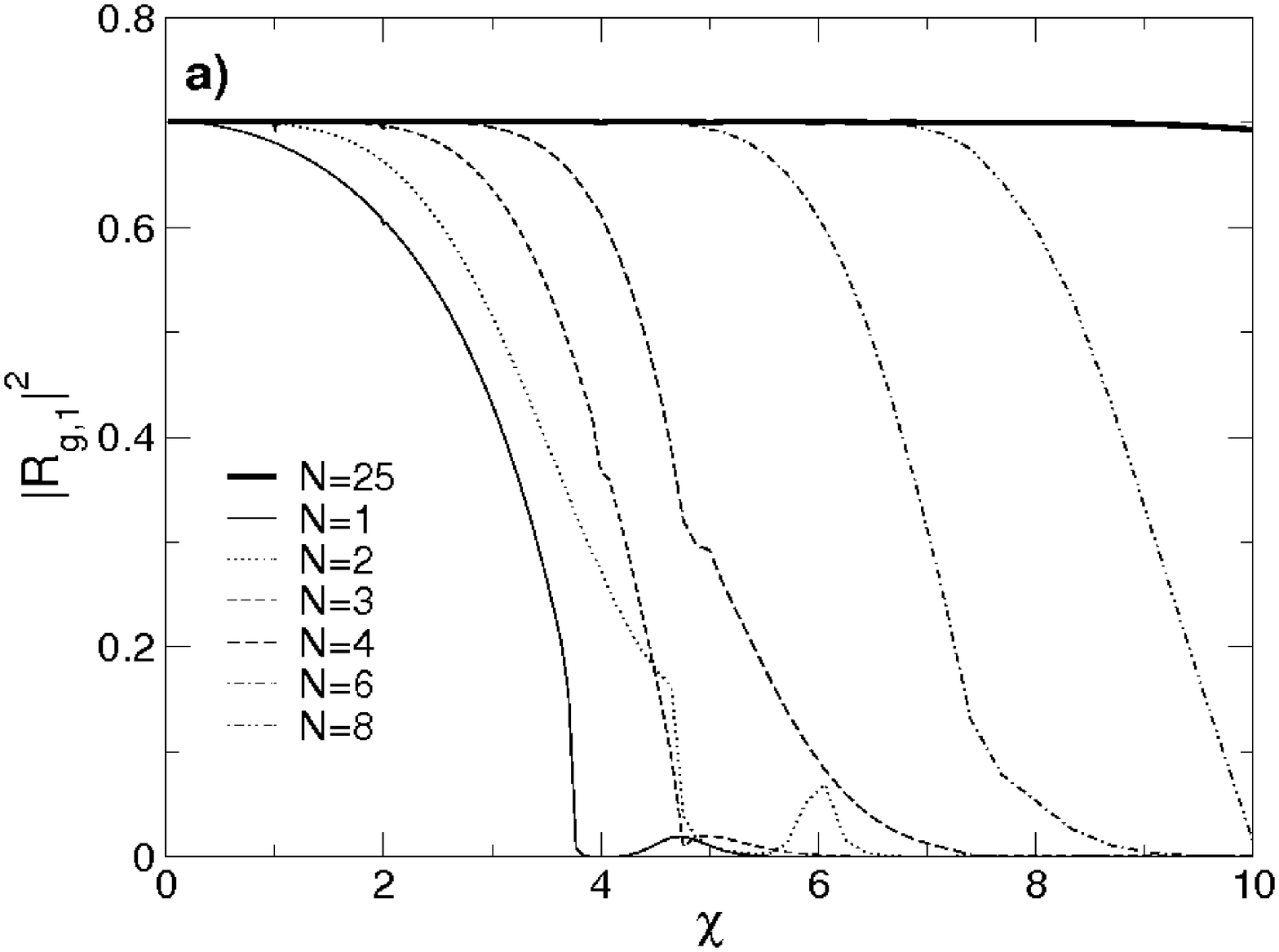}
\\
\vspace{1cm}
\includegraphics[height=6cm]{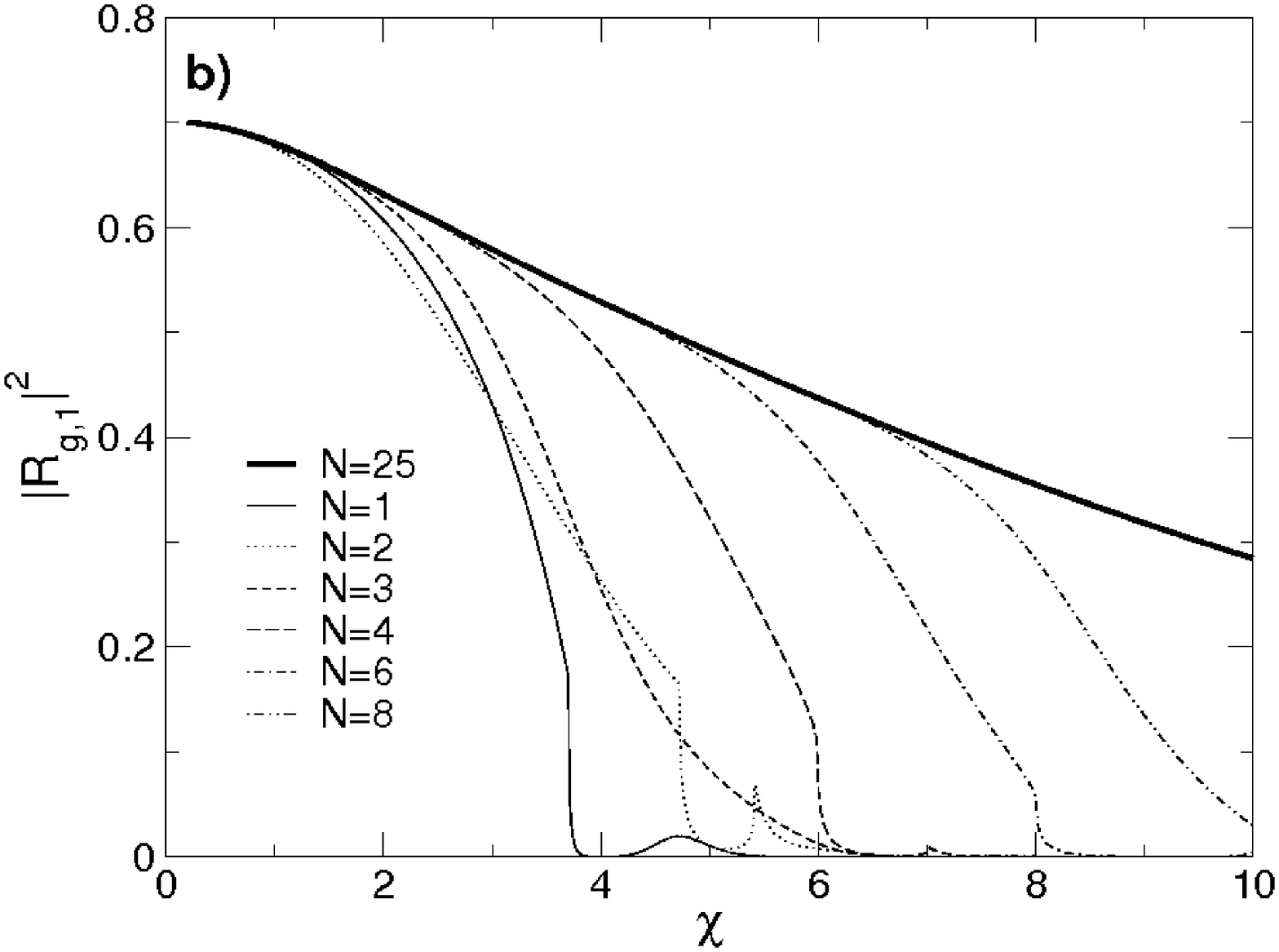}
\caption[]{Reflection in the ground internal and first transversal mode as a 
function of $\chi\equiv{k_L a}/{\pi}$ calculated for different $N$. Note that each line deviates from the 
exact one (thick line) around integer values of $\chi$.
The data are for a Ne atom with $v=0.1$ ms$^{-1}$ (which will be used in the rest of the paper)
and Rabi frequency $\Omega={10^8}$ s$^{-1}$. In (a) the waveguide width
is $a=1$ $\mu$m, and $a=100$ nm in (b).}
\label{convergence}
\end{figure}
\subsection{Convergence \label{convergence_section}}
We now study the convergence with respect to the number of 
free transversal modes $N$ included in the truncated basis. 
It turns out that the critical parameter is 
the ratio between the photon momentum $k_L\hbar$ and the quasi-momentum jump 
$\hbar\pi/a$, which determines the accessibility of transversal
energy levels in the waveguide by laser excitation. 
Intuitively, if the laser does not provide enough transversal momentum to connect 
the different quasi-momenta in the free modes of a given transition, the transition will not occur and may be ignored.    
As it can be seen in Fig. \ref{convergence}, each approximation for a fixed $N$ is
valid until $\chi$, ($\chi\equiv k_La/\pi)$, reaches an integer value. For example, the two-mode approximation
($N=2$)
will be valid as long as $\chi<1$, the $N=3$ model will be valid for
$\chi<2$, and so on.  In Fig. \ref{convergence}, the reflection probability   
in the ground state and first transversal mode, $|R_{g,1}|^2$, has been depicted, but a similar behavior is observed for other probabilities.
(For comparison with the language used in ion traps \cite{IT2,IT3},
note that $\chi=\eta/\pi$, where $\eta=k_L a$
is a Lamb-Dicke parameter.)     

It is remarkable that the validity of this convergence criterion is quite independent 
of the number of open channels, i.e., the number of modes that can be populated 
asymptotically in the free region based on energy considerations. 
In Figs. \ref{convergence}(a) and
\ref{convergence}(b) different waveguide widths are considered.
In Fig. \ref{convergence}(a), the waveguide width  is such that $10$ free transversal modes 
are open whereas in Fig. \ref{convergence}(b) only one mode is open. Comparing both figures, it is clear that the important parameter is $\chi$ and 
not the number of energetically open channels.

An exception of the above convergence rule occurs for values 
of $\Omega$ with an  
``avoided crossing'' between two $\epsilon_\alpha$ levels.  
This will be discussed in the following section. 
%
%
%
%
%
\section{Transversal excitation}
%
%
%
%
%
\begin{figure}
\includegraphics[height=6cm]{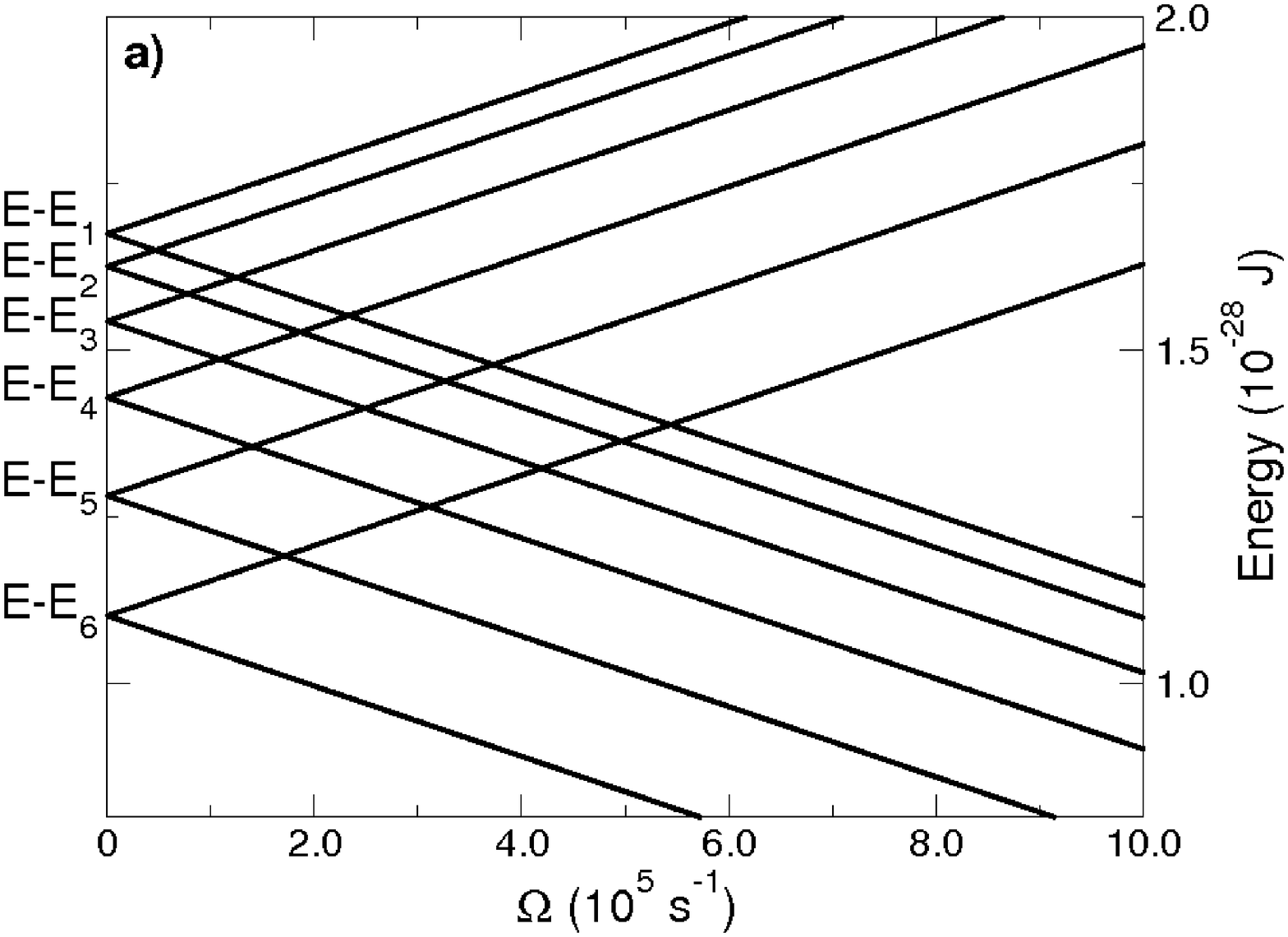}
\\
\vspace{1.2cm}
\includegraphics[height=6cm]{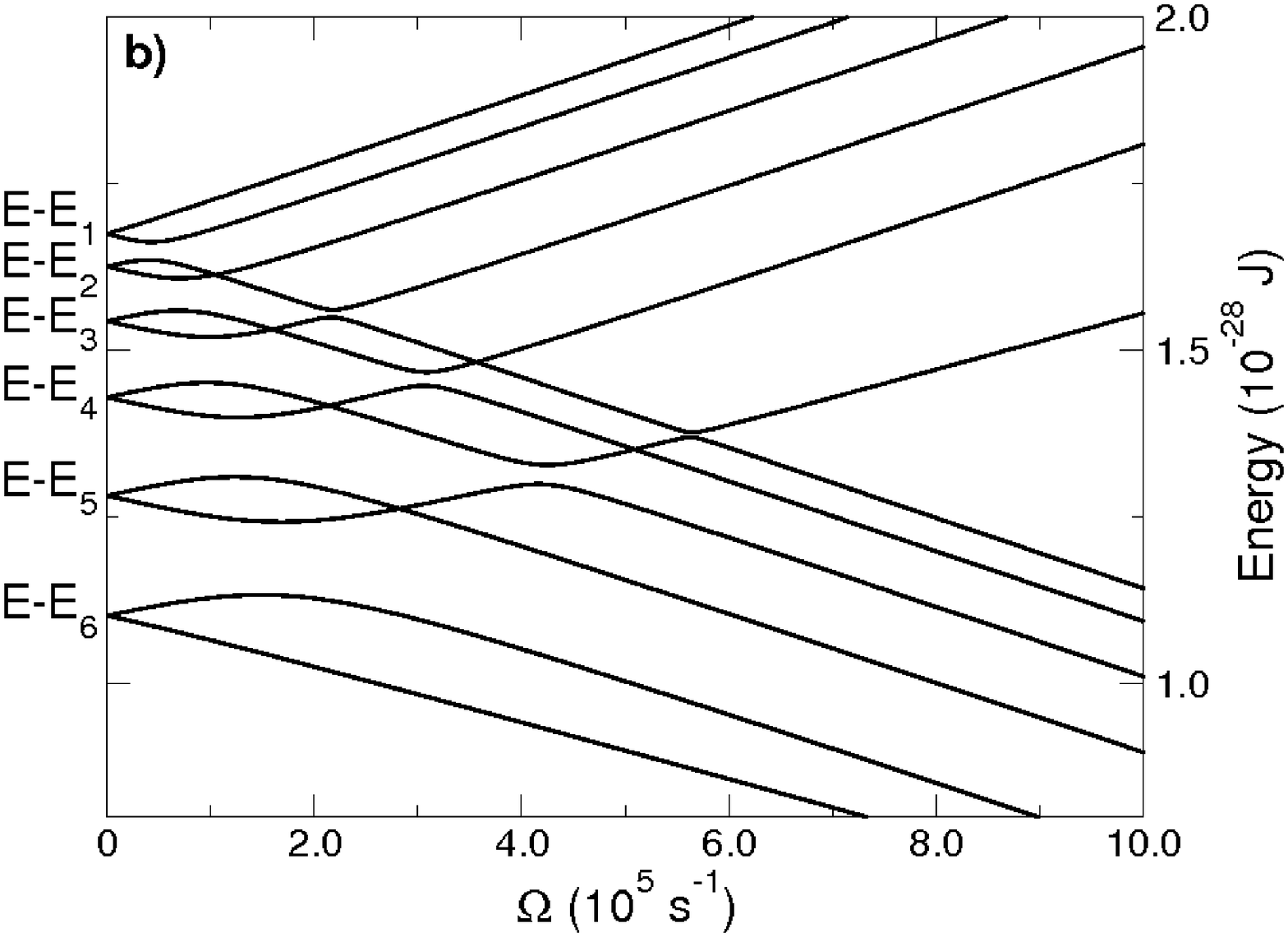}
\\
\vspace{1.2cm}
\includegraphics[height=6cm]{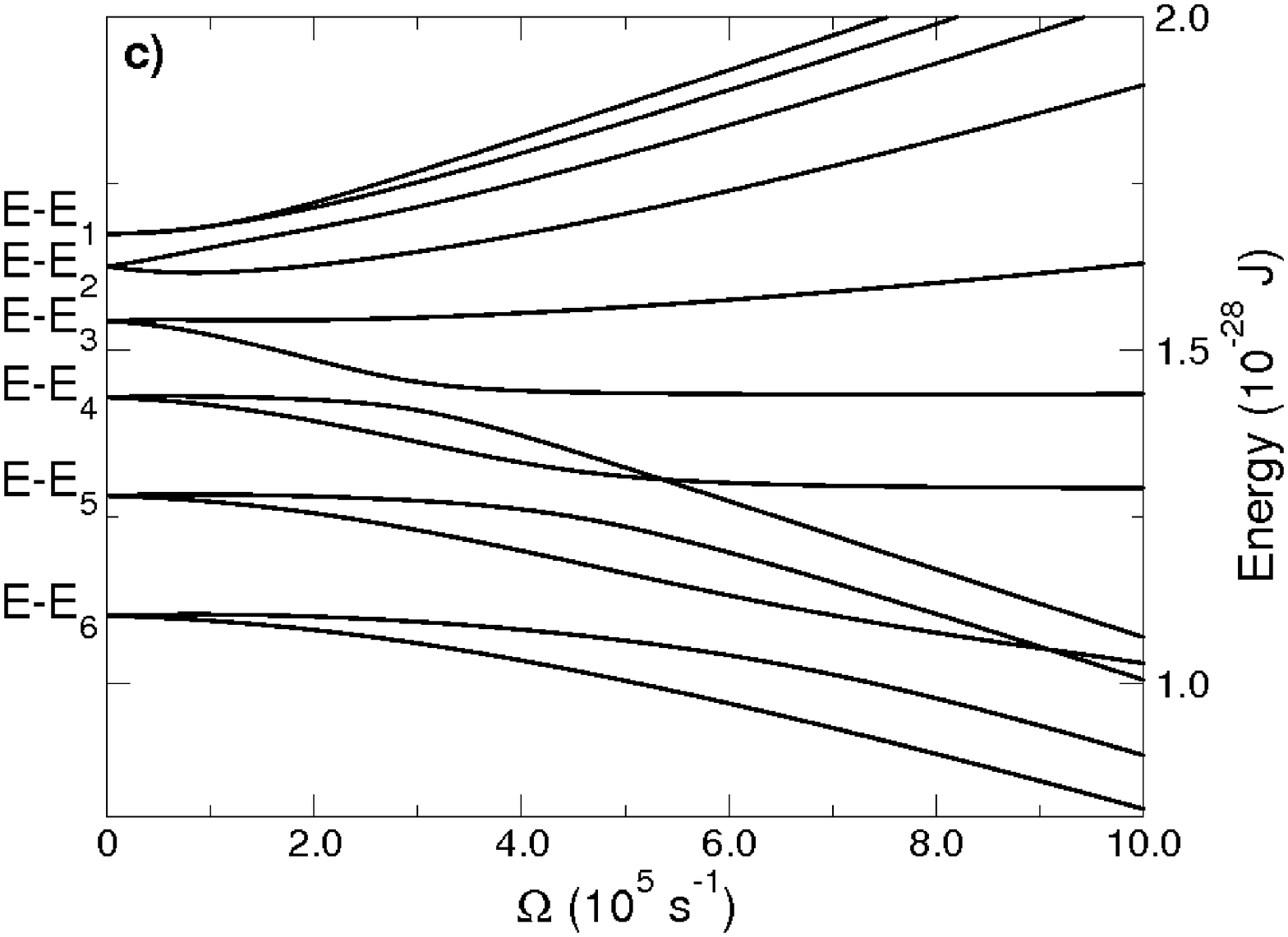}
\caption[]{Energy levels $\epsilon_\alpha$ inside the laser as a function of
the Rabi frequency for $N=6$ and $a=1$ $\mu$m:(a) No  
transversal coupling ($\chi = 0$);   
(b) Weak coupling, $\chi\simeq0.8$; (c) Strong coupling, 
$\chi\simeq4.3$.
Note that the rightmost crossing in (b) is avoided. For a Ne atom with $v=0.1$ ms$^{-1}$ the 
value of the total energy is $E=1.69\times10^{-28}J$.}
\label{eigenvalues}
\end{figure}
%
In this section we are examining the transmission probabilities for different values of $\chi$. 
The eigenvalues of $\hat {\cal{E}}$, 
Eq. (\ref{a_definition}),  are the longitudinal kinetic energies 
available for each mode in the laser region. In Fig. \ref{eigenvalues} 
typical diagrams of these
energy levels $\epsilon_\alpha$ as a function of the Rabi frequency are
plotted. Very different level structures are observed
depending on the transversal coupling factor $\chi = k_L a/\pi$.
In the no-coupling limit, Fig. \ref{eigenvalues}(a), 
the levels are straight lines and 
all the crossings are permitted. In the weak coupling case of Fig. \ref{eigenvalues}(b), 
the global structure of these energy levels 
is preserved except in some areas with avoided crossings. 
By 
increasing the coupling parameter further, the level structure becomes 
more and more distorted, see Fig. \ref{eigenvalues}(c).
We shall now study the dynamics inside the laser field in
these different regimes.
%
%
\subsection{No-coupling limit}
%
We consider first the limiting case $\chi = 0$ in which  
the coupling elements (\ref{coupling_elements_1}) are given by
$$
C_{nn'}^{\pm}=\delta_{nn'},
$$
%
so the transversal free modes do not couple. 
The $\hat {\cal{E}}$ matrix becomes 
\begin{equation}
\label{a_0}
\hat {\cal{E}}= \left( \begin{array}{c|c}
\begin{array}{ccc}
E-E_1&&\\
&\ddots&\\
&&E-E_N
\end{array}&-\frac{\hbar\Omega}{2}\boldsymbol{1_N}\\
\hline
-\frac{\hbar\Omega}{2}\boldsymbol{1_N}&
\begin{array}{ccc}
E-E_1&&\\
&\ddots&\\
&&E-E_N
\end{array}
\end{array}\right)
\end{equation}
where $\boldsymbol{1_N}$ represents the $N$-dimensional identity matrix.
The eigenvalues of $\hat {\cal{E}}$ are, as functions of $\Omega$,
the straight 
lines
\begin{equation}
\label{eigenvalues_0}
\epsilon_{n,\pm}^{(0)}=E-E_n\pm\frac{\hbar\Omega}{2},
\end{equation}
and the eigenvectors are labeled with the 
corresponding free mode $n$,
\begin{equation}
\label{eigenvectors_0}
|\epsilon_{n,\pm}^{(0)}\ra=\frac{1}{\sqrt 2} \left(
\mp|g,\varphi_n\ra +|e,\varphi_n\ra \right).
\end{equation}
Note that the absence of transversal coupling does not mean that the atom 
is uncoupled from the laser. The longitudinal velocity and internal states 
are affected and depend on $\Omega$.

\begin{figure}
\includegraphics[height=6cm]{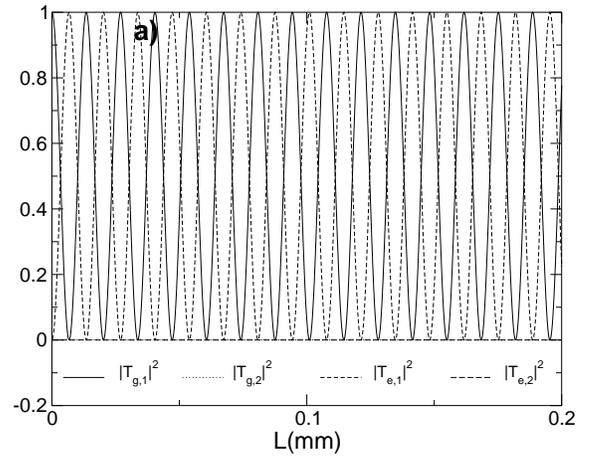}
\\
\vspace{1cm}
\includegraphics[height=6cm]{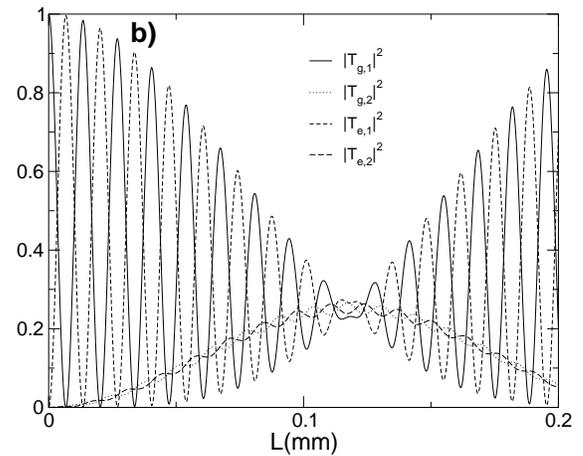}
\\\vspace{1.1cm}
\includegraphics[height=6cm]{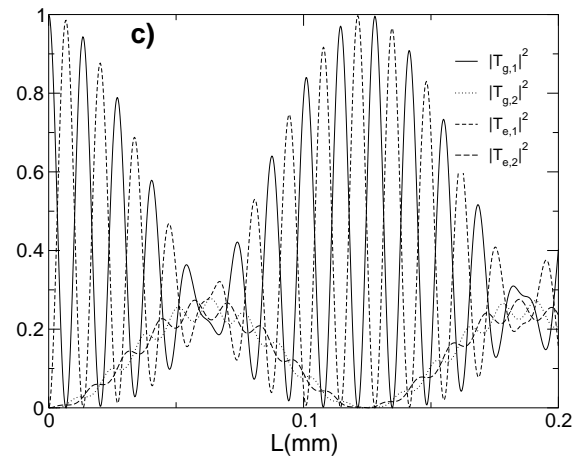}
\caption[]{Transmission probability vs. laser width $L$ 
for different values of $\chi$ around the crossing between 
$\epsilon_1$ and $\epsilon_2$ ($\Omega=\frac{E_2-E_1}{\hbar}\approx0.47\times 10^{5} s^{-1}$). 
In (a), $\chi = 0$;
in (b), $\chi \approx 0.3/\pi$,
and in (c) $\chi \approx 0.6/\pi$. In all cases  
$a=1$ $\mu$m.}
\label{rabi_osc}
\end{figure}

Now, we shall consider a semiclassical approximation valid for 
$ mv^2\gg\hbar\Omega $, so that the reflection is negligible,
and almost all atoms cross the laser region. In this approximation, 
from Eqs. (\ref{eigenvalues_0}) and (\ref{q_0}), the momentum inside the laser 
field may be approximated by
\begin{equation}
q_1^{\pm}\approx k_1 \left(1\pm \frac{\hbar\Omega}{2mv^2}\right).
\end{equation}
With the above eigenvectors, one can write from Eq. (\ref{gral_sol_x}) the
general wave function inside the laser.
Using the matching conditions and neglecting reflection, it can be shown that the only 
surviving eigenmode amplitudes inside the laser region are those related to 
the first longitudinal eigenmodes $|\epsilon_{1,\pm}\ra$ and are given by
\begin{eqnarray}
\label{popul1}
A_1^\pm&=&\frac{\mp 1}{2\sqrt 2} \left( 1+\frac{k_1}{q_1^\pm} \right),\\
\label{popul2}
B_1^\pm&=&\frac{\mp 1}{2\sqrt 2} \left( 1-\frac{k_1}{q_1^\pm} \right),
\end{eqnarray}
%
%
%
%
\begin{figure}
\includegraphics[height=6cm]{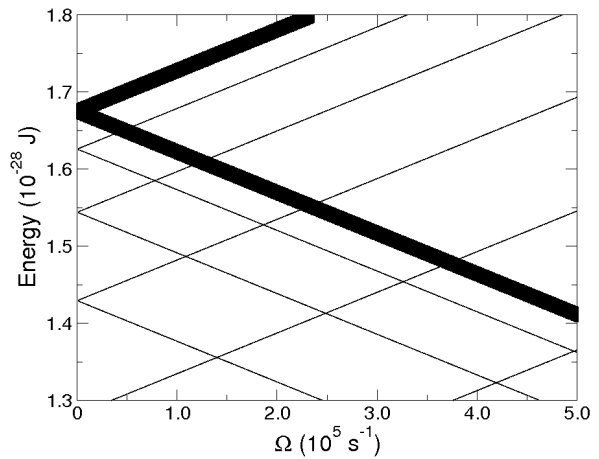}
\caption[]{Energy level diagram in the non-coupling limit $k_La/\pi=0$. 
The thick lines represent the populated modes and their thickness is proportional to the corresponding 
population, see Eq.  (\ref{probabilities}). A thickness of $0.04$ in the scale shown represents probability one, here 
and in Figs. \ref{weak_coupling} and \ref{strong_coupling}.
Only two states (those coming from $E-E_1$, associated with the ground transversal level
$n=1$) are populated.}
\label{no_coupling}
\end{figure}
where 
\begin{equation}
\label{q_0}
\left(q_n^\pm\right)^2={2m\epsilon_{n,\pm}^{(0)}}/{\hbar^2}.
\end{equation} 
According to Eqs. (\ref{popul1}) and (\ref{popul2}), only two states are populated, 
namely the states $|\epsilon_{1,\pm}^{(0)}\ra$,
both degenerate at $\Omega=0$, see Fig. \ref{no_coupling}.
These are combinations of ground and excited internal states multiplied by 
the transversal state $\varphi_1$, see Eq. (\ref{eigenvectors_0}). 
The populations of the different combinations of ground and excited internal and transversal states 
depend on $x$ as  (writing the dominant term in the semiclassical approximation for $0\le x\le L$)
\begin{eqnarray}
|\psi_{g,1} (x)|^2 &\approx& \cos^2\left[\left(\frac{q_1^+-q_1^-}{2}\right)x\right],
\nonumber\\
|\psi_{e,1} (x)|^2 &\approx& \sin^2\left[\left(\frac{q_1^+-q_1^-}{2}\right)x\right],
\nonumber\\
|\psi_{g,n} (x)|^2 &\approx& |\psi_{e,n} (x)|^2\approx0,\quad n=2,3,...
\label{ROS}
\end{eqnarray}
which are nothing but ``spatial Rabi oscillations''. 
The transmission probabilities are obtained by reading the populations at the edge 
of the laser region, i.e., replacig $x$ by $L$ in the last equation, see Fig. \ref{rabi_osc}(a)
\begin{eqnarray}
\label{transmission_1}
|T_{g,1}|^2 &\approx& \cos^2\left[\left(\frac{q_1^+-q_1^-}{2}\right)L\right],
\nonumber\\
|T_{e,1}|^2 &\approx& \sin^2\left[\left(\frac{q_1^+-q_1^-}{2}\right)L\right],
\nonumber\\
|T_{g,n}|^2 &\approx& |T_{e,n}|^2\approx0,\quad n=2,3,...
\label{ROS}
\end{eqnarray}
Note that, in the opposite limit of very slow incident atoms, i.e. $mv^2\ll\hbar\Omega$, the
phenomenon of Rabi Oscillation Suppression occurs as 
discussed in \cite{ROS}: the traveling plane waves inside the laser
become evanescent as the corresponding longitudinal wavenumber $q$ becomes
purely imaginary. This case will be ignored 
in the rest of the paper. 
%
%
%
\subsection{Weak coupling}
\begin{figure}
\includegraphics[height=6cm]{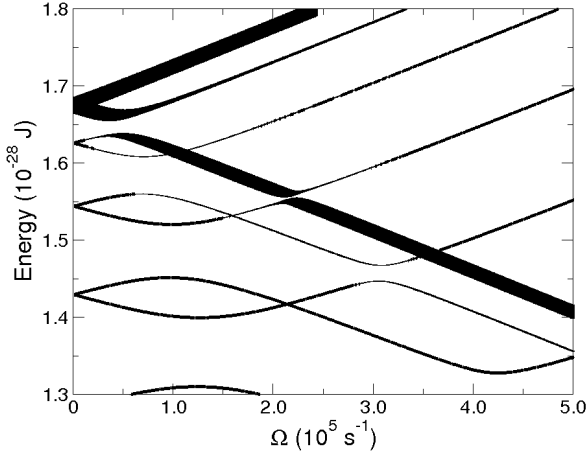}
\caption[]{Energy level diagram for $\chi\approx 3/\pi$ ($\lambda_L=2$ $\mu$m and $a=1$ $\mu$m),  
at the border of weak-coupling.  
The thickness of each line is again proportional to the population. 
Notice the alternation between avoided and permitted crossings and the population of 
excited transversal states at the avoided crossings.}
\label{weak_coupling}
\end{figure}

%
For small values of the transversal coupling parameter $\chi$, the coupling elements
(\ref{coupling_elements_1}) may be expanded in
power series of $\chi$. To first order in $\chi$, 
\begin{eqnarray}
\label{dnn_approx}
C_{nn}^{\pm}&\approx&1\pm i\pi\frac{\chi}{2}+\mbox{O}(\chi)^2,\\
C_{n\ne n'}^{\pm}&\approx&
\mp\frac{4nn'\left[1-(-1)^{n+n'}\right]}{(n^2-n'^2)^2\pi}i\chi
+ \mbox{O}(\chi)^2.
\label{dnn'_approx}
\end{eqnarray}
The $\hat {\cal{E}}$ matrix defined in (\ref{a_definition}) can then be divided as
$$
\hat {\cal{E}}=\hat{\cal{E}}_0+\hat V,
$$
%
where $\hat{\cal{E}}_0$ is the non-coupled $\hat {\cal{E}}$ matrix defined in (\ref{a_0}) and 
$\hat V$ is a small perturbation given by
\begin{eqnarray*}
\hat V&=&-\frac{\hbar\Omega}{2}\left( \begin{array}{c|c}
\boldsymbol{0_N}&\hat C ^{-}-\boldsymbol{1_N}\\
\hline
\hat C^{+}-\boldsymbol{1_N}&\boldsymbol{0_N}
\end{array}\right).
\end{eqnarray*}
The solutions to the non-perturbed part are already known from the previous section. 
Moreover, since 
\begin{eqnarray}
\la \epsilon_{n,+}^{(0)}|\hat V|\epsilon_{n,+}^{(0)}\ra&=&0,
\nonumber\\
\la \epsilon_{n,-}^{(0)}|\hat V|\epsilon_{n,-}^{(0)}\ra&=&0,
\nonumber
\end{eqnarray}
the energy levels have no first order corrections, i.e.,
$$
\epsilon_{n,\pm}\approx\epsilon_{n,\pm}^{(0)}
=E-E_n\pm\frac{\hbar\Omega}{2}.
$$
%
This perturbative method is valid except for those $\Omega$ near 
level-crossings. In these especial cases, 
a degenerate perturbation theory is required, but 
away from the crossings, the system behaves as in the non-coupled case 
described in 
the previous section:
Rabi Oscillations between internal states as those of Fig.
\ref{rabi_osc}(a)
are observed with no excitation of transversal modes,
see Fig. \ref{weak_coupling}.
%
\subsubsection{Permitted and avoided crossings}
To study the crossings between zeroth order levels 
$\epsilon_{n,\pm}^{(0)}$ and $\epsilon_{n',\mp}^{(0)}$
with different sign (otherwise no crossing may take place),  
a degenerate perturbation theory shall be used. The perturbation matrix 
$\hat V$ in the $2$-dimensional degenerate sub-space is
\begin{eqnarray}
\tilde V_{n,n'}&=&\la \epsilon_{n,\pm}^{(0)}|\hat V|\epsilon_{n'\mp}^{(0)}\ra
\nonumber\\
&=&\mp\frac{\hbar\Omega}{4}\left[ C_{nn'}^{+}-C_{nn'}^{-}\right],
\nonumber
\end{eqnarray}
where $n\ne n'$.
To first order $C_{nn'}^{+}=-C_{nn'}^{-}$, see Eq. 
(\ref{dnn'_approx}), and the relevant $2\times2$ 
potential matrix is 
\begin{equation}
\label{2times2_submatrix}
\tilde V_{[n,n']}=\frac{\hbar\Omega}{2}C_{nn'}^{+}
\left(
\begin{array}{cc}
0&1\\
-1&0
\end{array} \right).
\end{equation}
The eigenvalues, $\lambda_{\pm}=\pm i\frac{\hbar\Omega}{2}C_{nn'}^{+}$, 
provide the energy corrections to first order in $\chi$ to the 
energy levels. 
Using the approximate expression for $C_{nn'}^{+}$ in Eq. (\ref{dnn'_approx}), 
we obtain a general expression for the energy splitting at an arbitrary crossing, 
\begin{equation}
\label{energy_shift_gral}
\Delta\epsilon_{nn'}=\frac{8nn'\chi}{(n+n')^2(n-n')^2\pi}\left( \frac{\hbar\Omega}{2} \right)\left[1-(-1)^{n+n'}\right].
\end{equation}
Note that for even values of $n+n'$, $\Delta\epsilon_{nn'}=0$. In this case, no degeneration removal occurs 
and the crossing is permitted. This explains the alternation of avoided and permitted crossings 
in Figs. \ref{eigenvalues}(b) or \ref{weak_coupling}. 
%
%
\subsubsection{Transversal state excitation at avoided crossings: ``Rabi resonances''}
Near a Rabi frequency for which a crossing occurs, 
the states not participating in the crossing will still be given
by the combinations of free states in 
Eq. (\ref{eigenvectors_0}). Instead, the two levels taking part in the crossing 
are obtained from degenerate perturbation theory 
as the eigenvectors of the perturbation matrix (\ref{2times2_submatrix}).
Since we are concerned with incident atoms 
in the ground internal and transversal states, 
one of the levels involved will be $\epsilon_{1,-}^{(0)}$ and 
the other one $\epsilon_{n,+}^{(0)}$.  
The correct linear combinations of these states are 
\begin{eqnarray}
\label{combination}
|\epsilon_>\ra&=&\frac{1}{\sqrt{2}} \left( i|\epsilon_{1,-}^{(0)}\rangle
+ |\epsilon_{n,+}^{(0)}\rangle\right),\nonumber\\
|\epsilon_<\ra&=&\frac{1}{\sqrt{2}} \left( |\epsilon_{1,-}^{(0)}\rangle
+i |\epsilon_{n,+}^{(0)}\rangle\right),
\end{eqnarray}
where the subscripts $>$ and $<$ are related to the eigenvalues $\lambda_{\pm}$, i.e., 
upwards or downwards corrected levels.

In the semiclassical approximation with negligible reflection, the only 
non-zero eigenmode amplitudes are
\begin{eqnarray}
A_1^+=-\frac{1}{2 \sqrt 2} \left( 1+\frac{k_1}{q_1^+} \right)&,&\;
B_1^+=-\frac{1}{2 \sqrt 2} \left( 1-\frac{k_1}{q_1^+} \right),
\nonumber\\
A_>=-\frac{i}{4} \left( 1+\frac{k_1}{q_>} \right)&,&\;
B_>=-\frac{i}{4} \left( 1-\frac{k_1}{q_>} \right),
\nonumber\\
A_<=\frac{1}{4} \left( 1+\frac{k_1}{q_<} \right)&,&\;
B_<=\frac{1}{4} \left( 1-\frac{k_1}{q_<} \right).\nonumber
\end{eqnarray}
%
%
With the definitions of $|\epsilon_{n,\pm}^{(0)} \ra$ in
Eq. (\ref{eigenvectors_0}) and  
$|\epsilon_{>,<}\ra$ in Eq. (\ref{combination}), this will result in 
the following oscillations 
%
%
%
\begin{eqnarray}
|\psi_{g,1}(x)|^2&\approx&\frac{1}{2}\cos^2 \left[ \left( q_1^+-q_>\right)x\right] + 
\frac{1}{2}\cos^2 \left[ \left( q_1^+-q_<\right)x\right]\nonumber\\
&&-\frac{1}{4} \sin ^2\left[\left(q_>-q_<\right)x\right],
\nonumber\\
|\psi_{e,1}(x)|^2&\approx&\frac{1}{2}\sin^2 \left[ \left( q_1^+-q_>\right)x\right] + 
\frac{1}{2}\sin^2 \left[ \left( q_1^+-q_<\right)x\right]\nonumber\\
&&-\frac{1}{4} \sin ^2\left[\left(q_>-q_<\right)x\right],
\nonumber\\
|\psi_{g,n}(x)|^2&\approx&|\psi_{e,n}(x)|^2\approx\frac{1}{4}\sin^2\left[\left(q_>-q_<\right)x \right],
\end{eqnarray}
%
\begin{figure}
\includegraphics[height=6cm]{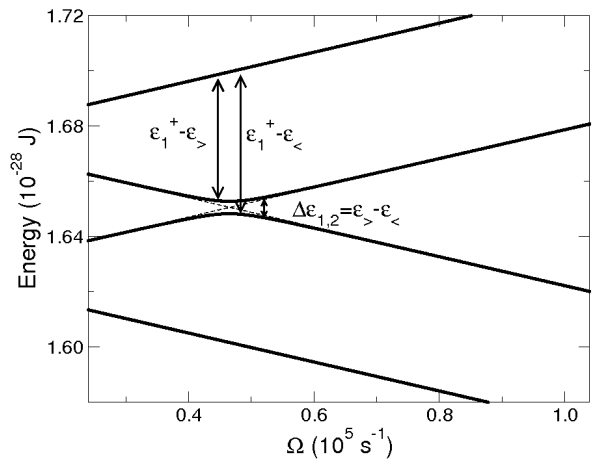}
\caption[]{Detail of the first avoided crossing ($\Delta\epsilon_{1,2}$)
in the weak-coupling case ($\chi\approx0.5/\pi$, 
with $a=1 \mu m$). The three different possible oscillation frequencies are marked by the arrows. The 
dashed lines represent the no-coupling $\chi=0$ limit. }
\vspace{1cm}
\label{crossing_zoom}
\end{figure}
%
from where it can be seen that the $n^{th}$ excited transversal state will 
only be populated whenever $q_{+}-q_{-}\ne0$, i.e., 
when the crossing is avoided. 
%
%
%
\subsubsection{Transmission}
The transmission amplitudes within the semiclassical
approximation are easily deduced from the form of the wave function at the edge 
of the laser region. In particular,  
for $\Omega$ 
around a Rabi Resonance, and in the weak coupling regime, $\chi\ll 1$, the transmission probabilities 
will be given by
\begin{eqnarray}
|T_{g,1}|^2&\approx&\frac{1}{2}\cos^2 \left( \frac{q_1^+-q_>}{2}L \right) + 
\frac{1}{2}\cos^2  \left( \frac{q_1^+-q_<}{2} L \right)\nonumber\\
&&-\frac{1}{4} \sin ^2 \left(\frac{q_>-q_<}{2}L \right),\nonumber\\
|T_{e,1}|^2&\approx&\frac{1}{2}\sin^2 \left( \frac{q_1^+-q_>}{2}L \right) + 
\frac{1}{2}\sin^2  \left( \frac{q_1^+-q_<}{2} L \right)\nonumber\\
&&-\frac{1}{4} \sin ^2 \left(\frac{q_>-q_<}{2}L \right),\nonumber\\
|T_{g,n}|^2&\approx&|T_{e,n}|^2\approx\frac{1}{4}\sin^2 \left(\frac{q_>-q_<}{2}L\right).
\end{eqnarray}
There are in principle three different oscillation
periods, associated with three energy splittings 
which can be identified around the crossing in Fig. \ref{crossing_zoom}.
However, for a narrow, sharp crossing
(i.e. for small splitting, $q_{>}-q_{<}\ll q_1^{+}-q_{>}\approx q_1^{+}-q_{<}$),
two of them become approximately equal so that 
essentially two main oscillation scales, a short and a long one, 
are identified in Figs. \ref{rabi_osc}(b,c).  In the no-coupling ($\chi=0$) limit
the crossing is permitted, then 
$q_{>}=q_{<}$, no transversal excitation occurs, and the standard Rabi 
oscillations between atomic internal states are found,
as in Fig. \ref{rabi_osc}(a).
%
Thus, the change of laser width $L$ gives us the opportunity to control the transmitted atomic internal 
and transversal state populations.
\subsection{Strong Coupling}
%
\begin{figure}
\includegraphics[height=6cm]{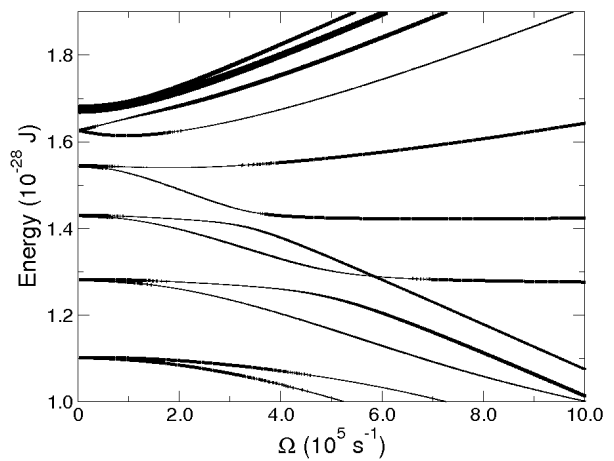}
\caption[]{Relative population of the levels in the strong coupling case
($\chi\approx13/\pi, a=1 \mu m, N=6$) 
represented by the line thickness.}
\label{strong_coupling}
\end{figure}
When the transversal coupling is stronger, the perturbative method is not valid anymore since the 
longitudinal energy levels inside the laser in Fig. \ref{eigenvalues}(c) are not small deviations 
from the straight lines in Fig. \ref{eigenvalues}(a). To find these levels 
the complete $\hat {\cal{E}}$ matrix defined in (\ref{a_definition}) has to be diagonalized. 

The behavior of the levels and their classification is not as simple as for weak coupling 
but some general trends may be observed, such as the allowed crossings between 
states which are degenerate at $\Omega=0$, or the flatness of the curves for small $\Omega$, 
due to the small values of the diagonal elements of the coupling matrix $\hat{C}$. 
Fig. \ref{strong_coupling} shows that the 
population remains mostly in the two upper levels. 
The avoided crossings are not as sharp and well defined as in the weak coupling case and, since  
the energy splitting is quite large, the ``population transfer'' from level to level is much less efficient.
\section{Conclusions and discussion}
We have investigated the effect of the interaction between 
two-level atoms in a rectangular, hard-wall waveguide and a laser with a square 
intensity profile, on resonance with an atomic transition, shining
perpendicularly to the waveguide axis. 
Since strong confinement and single mode dynamics are usually 
preferred, the weak transversal coupling case has been studied in detail, for  atom 
incidence in the ground internal and transversal states, but the results obtained  may be easily 
generalized for incidence in an arbitrary transversal or internal state, 
which is of interest if transversal cooling is intended. 
It is found that 
transversal excitation of the atom occurs only at avoided crossings in which the Rabi frequency 
coincides with the transition frequency between the ground and even transversal levels. 
The level splitting is proportional to the Rabi frequency and Lamb-Dicke parameter  $\eta=k_L a$, and also inversely 
proportional to the square of the  transition frequency for the transversal excitation. 
At sharp avoided crossings the atomic populations oscillate in the laser region (in space for stationary 
waves or in time for quasi-monochromatic wave packets) with a double frequency pattern. For certain laser widths 
that depend on the level splitting,  there results a transmitted wave with a homogeneous distribution 
among the four possible states combining ground and even excited
transversal states, and ground and excited internal states.
Other final states may be achieved by playing with laser detuning.  
Note that the transitions at transversal Rabi resonances are missed in the usual 
treatments of vibrational excitation in ion traps \cite{IT2,IT3}.
In that field the emphasis is on the effect of detuning, which has been taken as zero in the present work. Moreover, in that context
a second interaction picture is performed with respect to the vibrational Hamiltonian 
so that transitions such as the ones considered here, with  
involve excitation of both internal and transversal degrees of freedom
with zero detuning, are beyond the second rotating-wave approximation applied there, usually within the Lamb-Dicke regime
(equivalent to our weak coupling case, $\eta=k_L a<<1$).  
In future work, some of the idealizations of the present model will be removed, 
considering in particular harmonic rather than hard-wall confinement, detuning, and decay. 
Detuning combined with laser intensity and beam width are thus expected to serve as useful control knobs to promote or hinder 
transversal excitation or deexcitation and state selection, providing a promising route for optical atomic control in waveguides.  
%
%
%
%
\begin{acknowledgments}
We thank David Gu\'ery-Odelin for encouragement and useful comments.
This work has been supported by Ministerio de Educaci\'on y Ciencia
(BFM2003-01003),
and 
UPV-EHU (00039.310-15968/2004).\\
\end{acknowledgments}
%
%
%
%
%

\end{document}